\documentstyle[aps,prl,multicol]{revtex}

\begin{document}
\title{Uniqueness of Entanglement Measure and Thermodynamics}
\author{Vlatko Vedral and Elham Kashefi}
\address{Optics Section, The Blackett Laboratory,
Imperial College, Prince Consort Road, London SW7 2BW, England}
\date{\today}
\maketitle

\begin{abstract}
We apply the axiomatic approach to thermodynamics presented by Giles to derive a
unique measure of entanglement for bi-partite pure states. This implies that local
manipulations of entanglement in quantum information theory and adiabatic
transformations of states in thermodynamics have the same underlying mathematical
structure. We discuss possible extensions of our results to mixed and multi-partite
states.
\end{abstract}

\begin{multicols}{2}

Understanding of entanglement and its characterization form the cornerstone of the new
and rapidly growing filed of quantum information and computation \cite{Vrev}. We need
to know how much entanglement is at our disposal since entanglement is a form of
resource that can enhance information processing \cite{PV98}. Although a great deal of
work has recently been performed in this direction \cite{Hrev}, it is widely
acknowledged that we do not have a complete understanding of even the bi-partite
entanglement for mixed states. There is a number of measures to quantify entanglement
which apply in different settings and have different properties \cite{Hrev}. The
consensus, however, is that local operations aided with classical communication (LOCC)
are the key to explaining entanglement \cite{BBP96,PR97,VP98}. They are able to
separate disentangled states from entangled states and thus introduce a {\em
directionality} to entanglement manipulation processes: an entangled state can always
be converted to a disentangled one by LOCC, but not vice versa.

A comparison with thermodynamics will be very helpful at this point. The Second Law of
thermodynamics tells us which (energy conserving) processes are allowed in nature {\em
without any reference to the underlying physical structure}. The central role is
played by adiabatic processes and entropy is used to separate the possible from the
impossible processes according to a very simple principle: if a state $A$ has more
entropy than $B$, then there is an adiabatic process to go from $B$ to $A$, but not
vice verse. This was first clearly realized by Caratheodory \cite{Cara09}, who
restated the Second Law by saying that in the neighborhood of any state there exist
states which are adiabatically inaccessible from it. This allowed him to derive an
entropy function which is able to introduce ordering into the set of physical states.
The Second Law thus tells us that adiabatic processes cannot decrease entropy of the
system itself. The question, then, is whether entropy is the only such function. To
answer this question, however, thermodynamics needed first to be put onto a more
secure mathematical foundation. In the words of Caratheodory himself: "What
Thermodynamics needs is the establishment of logical order, essentially an
intellectual cleanup. This is a problem for a mathematician. The fundamental ideas and
concepts have been introduced by physicists long ago and a mathematician need not
worry about it". The first such formalization came from Giles in $1964$ \cite{Giles}
and was recently extended by Lieb and Yngvson \cite{LY98}.

An ideal physical theory should consist of two independent parts: a mathematical
theory and a set of rules of interpretation of various mathematical objects involved
in the theory. By formalizing a physical theory in such a way as to divorce it
completely from the physical interpretation, it is possible to derive a mathematical
structure that may be useful in a completely different physical setting to the
original one. The question we would like to address in this letter is whether it is
possible to apply the thermodynamical reasoning of Giles and Lieb and Yngvson to
entanglement manipulations to derive a unique measure of the amount of entanglement in
a general setting. The analogy between thermodynamics and quantum information would
identify adiabatic processes with LOCCs and entropy with entanglement. The aim is to
order quantum states according to whether they can be converted into each other by
LOCCs. We use Giles' approach to show that the same abstract structure of
thermodynamics captures entanglement manipulations in quantum mechanics. As a
consequence we can derive a rigorous proof for the unique measure of entanglement for
pure states in the same way that a unique measure of order (entropy) is constructed in
thermodynamics. We note that Giles' approach is more suitable for entanglement
manipulations than the approach of Lieb and Yngvson. This is because, as we will see
later, Giles bases his approach on transformation of a discrete number copies of a
physical state, while Lieb and Yngvson rephrase the axioms in terms of continuous
mixtures of states. The latter could also be applied to entanglement, but it is in our
view not as naturally suited as Giles' treatment.

Our abstract approach to entanglement is different to the existing method where one
looks for a minimal number of conditions for a measure of entanglement that would
single out a unique one \cite{VP98,Hor01,Rud01,DHR01}. The existing method has a
strong flavor of Shannon's pioneering approach to information theory \cite{Shannon}.
Shannon considered functions on the set of probability distribution which would
describe their information content. By introducing three natural conditions that this
function should satisfy he arrived at a unique measure, called the Shannon entropy.
These conditions are remarkably similar to the conditions leading to a unique measure
of entanglement for pure bi-partite states \cite{Rud01}. This, of course, is not
surprising. It is well known that the Shannon entropy of the probabilities derived
from Schmidt coefficients in the Schmidt decomposition \cite{Schmidt} of a pure
bi-partite state is a good measure of entanglement \cite{BBP96}.

The remaining part of the letter is structured as follows. First we introduce the
formal theory of Giles and we list the set of axioms he introduced to capture
thermodynamical processes. Secondly we define the notion of LOCC which
allows us to apply the presented axioms to entanglement manipulations. Finally, we
compare our approach to the other existing approach and discuss the domain of its
applicability.

To begin, we study a non-empty set $\cal{S}$, whose elements are called {\it states},
in which two operations, $+$ and $\rightarrow$, are defined. States are denoted by :
$a,b,c,\dots$ . A {\it process} is an ordered pair of states $(a,b)$. In the following
axioms we will omit the phrase "for all ...".

\vspace{0.2cm}
\noindent{\bf Axioms 1-5.}
\begin{itemize}
\item[1.] {\em The operation $+$ is associative and commutative};
\item[2.] $a \rightarrow a$;
\item[3.] $a \rightarrow b \;\; \& \;\; b \rightarrow c \Longrightarrow a \rightarrow c$;
\item[4.] $a \rightarrow b \Longleftrightarrow a+c \rightarrow b+c$;
\item[5.] $a \rightarrow b \;\; \& \;\; a \rightarrow c \Longrightarrow b \rightarrow c
 \;\; \mbox{{\it or}} \;\; c \rightarrow b$;
\end{itemize}
Let us briefly discuss why these axioms describe the structure of thermodynamics. The
operation $+$ represents the physical operation of considering two systems together.
Therefore it must naturally be associative and commutative. The operation
$\rightarrow$ represents an adiabatic process which is meant to convert different
physical states into each other. Therefore, like any other physical process, it should
naturally be reflective and transitive as in axioms $2$ and $3$. Axiom $4$ is the
first non-intuitive property linking the operations $+$ and $\rightarrow$. In the
forward direction it is obvious that if state $a$ can be converted into $b$ then the
presence of another state $c$ should not alter this fact, i.e. we can convert $a$ and
$c$ into $b$ and $c$ by converting $a$ into $b$ while doing nothing to $c$. In the
backward direction, however, this axiom is not completely obvious. It says that if a
process is possible with the aid of another state, then we, in fact, do not need this
other state for the process. Thermodynamics deals with macroscopic systems with a
large number of degrees of freedom (subsystems). It is in the "asymptotic" limit that
this axiom becomes more natural. Finally, axiom $5$ is the key property which allows
us to compare different states and processes. It says that any two states that are
accessible from a third states must be accessible to each other at least in one
direction. Not being able to do so would lead to states which would be incomparable as
there would be no physical way of connecting them. Thus a unique way of ordering
states would be impossible.

We now need the following definition in order to introduce the remaining two axioms.
It is important for comparing the sizes (contents/amounts) of different physical
states.

\vspace{0.2cm} {\bf Definition.} Given states $a$ and $b$ we write $a \subset b$ ({\it
and say that $a$ is contained in $b$}), if there exists a positive integer $n$ and a
state $c$ such that
\begin{eqnarray*}
na+c \rightarrow nb \;\; \mbox{or} \;\; nb \rightarrow na+c
 \;\; \mbox{.}
\end{eqnarray*}
This really says that the state $a$ is smaller than $b$ if $a$ requires the help of
another state $c$ to be converted to or derived from $b$.

\vspace{0.2cm} {\bf Definition.} A state $e$ is an {\it internal} state if, given any
state $x$, there exists a positive integer $n$ such that $x \subset ne$.
\vspace{0.1cm}

\noindent This definition serves to introduce a reference state, which is the one that
can contain any other physical state given sufficiently many copies of it. The concept
of the internal state is necessary to give a basic metric unit (yardstick) to quantify
the physical content of a state in a unique way (i.e. independent of states).

Now we state the remaining two axioms.

\vspace{0.1cm} \noindent{\bf Axioms 6-7}
\begin{itemize}
\item[6.] {\it There exists an internal state};
\item[7.] {\it Given a process $(a,b)$, if there exists a state $c$ such that for any positive real number
$\epsilon$ there exists positive integers $m$, $n$ and states $x$, $y$ such that $m/n
< \epsilon$, $x \subset mc$, $y \subset mc$, and $na+x \rightarrow nb+y$ then $a
\rightarrow b$};
\end{itemize}
\vspace{0.1cm} Axiom $6$ is necessary if we are to compare contents of different
states in a unique way. Axiom $7$ is the most complex axiom in the theory, although it
is strongly motivated by the logic of thermodynamical reasoning. Loosely speaking, it
states that if we can transform $a$ into $b$ with an arbitrarily small environmental
influence, then this influence can be ignored. This, in some sense, introduces
continuity into thermodynamical properties (and will be crucial for entanglement
manipulations later). A reader interested in a more detailed physical interpretation
of the axioms is advised to consult Giles' book \cite{Giles}.

The above $7$ axioms are the crux of Giles' formal theory which capture the key idea
behind the Second Law of thermodynamics. Based on them Giles proves the existence of
the unique {\it entropy function}. However, his formalization is general enough to be
applicable to local manipulations of quantum states leading to a rigorous proof of the
existence of the unique entanglement function for pure states.

Now to apply the axioms to entanglement manipulations consider $\cal{S}$ to be the set
of all quantum pure states and the operation $+$ to be the tensor product $\otimes$.
In order to define the operation $\rightarrow$, we first give the definition of
$\subset$ in the quantum setting and then extend the concept of LOCCs in the spirit of
axiom $7$.

\vspace{0.1cm} {\bf Definition.} We say that a pure state $a$ is contained in a pure
state $b$, designated as $a \subset b$, iff there exists an integer $n$ and a state
$c$ such that either of the following two cases is valid
\begin{eqnarray*}
&\mbox{i)}&\;\;(\forall \epsilon)(\exists \Phi \in \mbox{\bf LOCC}): ||\Phi(a^{\otimes n}\otimes c) -
 b^{\otimes n}||<\epsilon
 \\
&\mbox{ii) }&\;\;(\forall \epsilon)(\exists \Phi \in \mbox{\bf LOCC}): ||\Phi(b^{\otimes n}) -
 a^{\otimes n}\otimes c||<\epsilon
 \;\; \mbox{.}
\end{eqnarray*}
Thus we say that a quantum state $a$ is contained within a state $b$ if, with the help
of some other state $c$, $a$ can be transformed by LOCC into $b$. Now we define what
we mean by a transformation of one quantum state into another.

\vspace{0.1cm} {\bf Definition.} We say that a pure state $a$ can be converted into a
pure state $b$, designated as $a \rightarrow b$, iff
\begin{eqnarray*}
&&(\forall \epsilon)(\exists c)(\forall \delta )(\exists  n,m\in N, \;\;\Phi\in
 \mbox{\bf LOCC} \;\; \mbox{and} \;\; x,y \in {\cal{S}}) \\
&&\mbox{ such that}\;\; m/n < \delta \;\; , \;\; x\subset mc \;\;,\;\;y \subset
 mc \;\; \mbox{and} \\
&&
||\Phi(a^{\otimes n}\otimes x) -
  b^{\otimes n}\otimes y||< \epsilon
 \;\; \mbox{.}
\end{eqnarray*}
Note that we define arrow such that axiom $7$ is immediately satisfied. This is
motivated by Giles' suggestion \cite{Giles} who regarded this axiom as a more general
definition of a process which captures real physical processes more precisely. We can
see two important points in applying Giles' formalism to entanglement manipulations.
One is that the approach is naturally asymptotic. This is of no surprise if we expect
a unique measure of entanglement to emerge. In fact, if we deal with a finite number
of copies \cite{Niel99}, then we know that axiom $4$ does not hold in the backward
direction. Namely, we can have a situation where $a$ cannot be transformed into $b$,
but that this could be achieved with a presence of an additional state $c$ (known as
the catalyzer \cite{JP99}). Point two is that our definition of LOCC is not only
asymptotic but allows the presence of an additional system as long as the state
returned at the end of the process is included in the same state as the original
\cite{Vedral99}. Therefore our definition of LOCC is tailored so that it naturally
satisfies all the axioms of Giles. What remains to be shown is that these axioms lead
to a unique measure of entanglement imposed on pure quantum states. Following Giles'
approach we first need to define the notion of the "component of content function".
This notion is important as the resulting entanglement measure will be unique up to
the addition of this function.

\vspace{0.1cm} {\bf Definition.} A real-valued function $Q$ defined for every state is
a {\it component of content function} if
\begin{itemize}
\item $Q(a\otimes b)= Q(a) + Q(b)$;
\item $a\rightarrow b \Longrightarrow Q(a)=Q(b)$;
\end{itemize}
By entanglement measure we will mean the following:

\vspace{0.1cm} {\bf Definition.} A real-valued function $E$ defined for every state is
a {\it entanglement function} if
\begin{itemize}
\item $E(a\otimes b)=E(a) + E(b)$;
\item If $a\rightarrow b \;\; \& \;\; b\rightarrow a$, then $E(a)=E(b)$;
\item If $a\rightarrow b \;\; \& \;\; b\rightarrow a$, then $E(a)>E(b)$;
\end{itemize}
\vspace{0.1cm} The following two theorems, stated here without proof, characterize the
entanglement function.

\vspace{0.1cm} {\bf Theorem 1.} Let $S_1$ be an entanglement function. If $Q$ is a
component of content and $\lambda$ a positive real number then $\lambda S_1 + Q$ is an
entanglement function; moreover, any entanglement function $S$ may be written in this
form.

\vspace{0.1cm} {\bf Theorem 2.} There exists a positive entanglement function.
\vspace{0.1cm}

The proof is constructive and can be found in \cite{Giles}. It leads to a definition
for a measure of entanglement, which in context of theorem $1$ is unique. When applied
to entanglement manipulations Giles' definition becomes simpler. The basic reason for
this is that our definition of arrow satisfies a stronger condition than stated in
axiom $5$. Namely, given any two pure bi-partite states $b$ and $c$, we have that
either $b\rightarrow c$ or $c \rightarrow b$. The definition of the amount of
entanglement of $a$ now becomes

\vspace{0.1cm}
\[
E(a)= \inf \{m/n : y\subset e^{\otimes m} y \rightarrow a^{\otimes
 n}\otimes x \}
 \;\; \mbox{,}\]
where the infimum is taken over all integers $m,n$ and all states $x,y$ and $e$ is any
internal state. This definition of entanglement has strong resemblance to the notions
of entanglement of formation \cite{BDS96}. Indeed, suppose that the state $e$ is some
maximally entangled state of two qubits. Then the above measure of entanglement of a
state $a$ looks at the (asymptotic) minimal number of copies, $m$, of the state $e$ we
need to invest to obtain $n$ copies of $a$, by LOCC with the aid of a catalyzer. Note
that we may well define the amount of entanglement of a state $a$ to be the maximum
number $m$ of maximally entangled states we can obtain from $n$ copies of $a$ when $n$
is large. In this case we have the entanglement of distillation (interestingly, this
is a concept that Giles does not consider in thermodynamics). Otherwise, we can use
any other internal state $e$ (e.g. a non-maximally entangled state) and all the
resulting measures of entanglement would be equivalent up to an affine transformation.

We now compare our approach with the other existing approach for the uniqueness of
entanglement measure for pure states due to Donald et. al. \cite{DHR01}. Building on
the results of Popescu and Rohrlich \cite{PR97} and Vidal \cite{Vidal}, they show that
there are three conditions necessary for the existence of a unique entanglement
measure for pure states: 1) Additivity, 2) Monotonicity and 3) Asymptotic continuity.
The unique measure resulting from these three axioms is equal to the von-Neumann
entropy of entanglement. Note that if in our definition of entanglement $e$ is taken
to be a maximally entangled state, then the amount of entanglement becomes $E(a) =
\inf \{m/n : e^{\otimes m} \rightarrow a^{\otimes n}\otimes x \}$. It is very simple
to show that this quantity is equal to the von Neumann's entropy of entanglement. This
therefore offers a different way of proving that Giles' axioms lead to a unique
measure of entanglement.

In Giles' formalism there is no reference to the number of subsystems or their
dimensionality involved in entanglement manipulations. Thus, it may be tempting to
conclude that there is a unique measure of entanglement for pure states of multi-party
systems. However, it is in general not clear that the axiom $5$ will hold for more
than two subsystems. For example, suppose that the state $b$ is a tri-partite state
$|\psi_1\rangle\otimes (|0_20_3\rangle + |1_21_3\rangle)$ while the state $c$ is
$(|0_10_2\rangle + |1_11_2\rangle)\otimes|\psi_3\rangle$. Then these states cannot be
converted into each other by our definition of $\rightarrow$ (where each party $1,2,3$
acts locally with the help of catalyzers and is allowed to classically communicate to
other parties) even though both can be derived from a common state $a$ of the form
$|0_10_20_3\rangle + |1_11_21_3\rangle$. This suggests that either the notion of a
unique measure of entanglement is misplaced for more than two parties or that we have
to extend the notion of $\rightarrow$ to satisfy axiom $4$. The latter is, however,
difficult to imagine as we are already allowing quite a wide range of local
operations.

The situation is somewhat similar in the bi-partite mixed state case. The axiom $5$
may not be satisfied in general although there is no proof of this. If the axiom $5$
fails, this would imply that there are states $b$ and $c$ which cannot be compared by
converting them into each other even though they can be derived from a common source.
This in turn means that arbitrary order may be assigned to them, i.e. according to one
measure $b$ would be more entangled than $c$ and according to some other measure it
could be the other way round without contradicting other axioms. Virmani and Plenio
\cite{VP00} have already shown that if two measure of entanglement coincide on pure
states and order mixed states differently, then they must be different measures (i.e.
they cannot be equivalent up to an affine transformation). Therefore we may again be
in a situation that if we want to have a unique measure of entanglement we need to
extend the definition of $\rightarrow$. A possible avenue to explore is to consider
PPT processes defined by Rains \cite{Rains}.

In this letter we have shown that the set of axioms originally presented by Giles to
capture the Second Law of thermodynamics can be applied to entanglement manipulations.
This leads to a unique measure of entanglement for pure states in the same way that
Giles showed the existence of a unique measure of order in thermodynamics. In order to
satisfy the axioms we had to define the notion of local operations to include the
possibility of catalyzers. Once catalyzers and arbitrarily large numbers of copies are
allowed in manipulations of entanglement we see that Giles' axioms are naturally
satisfied. This then establishes the fact that not only are there analogies between
thermodynamics and entanglement \cite{PR97,VP98,HHH98}, but that the two, in fact,
have {\em exactly} the same underlying structure. Therefore, it is of no surprise
that same entropic measure is used to order both thermodynamical states as well as
quantum entangled states. Our work suggests that this approach may be extended to
consider mixed and multi-party quantum states.

\noindent {\bf Acknowledgments.} We thank J. Eisert, M. Plenio and G. Vidal for useful
discussion on the subject of this paper. This work is funded by EPSRC and the European
grant EQUIP (contract IST-1999-11053).

\end{multicols}
\end{document}